\documentclass[twocolumn,aps,prl,superscriptaddress,longbibliography]{revtex4-1}
\usepackage[colorlinks,bookmarks=false,citecolor=blue,linkcolor=red,urlcolor=blue]{hyperref}
\usepackage{verbatim}
\usepackage{color}
\usepackage{amsmath,amssymb,bm,braket,float,mathtools,subfigure}
\usepackage{graphicx,xfrac,appendix}
\usepackage[english]{babel}

\makeatletter

\usepackage{epstopdf,float}

\makeatother

\begin{document}

\title{Exploring Metallic-Insulating Transition and Thermodynamic Applications of Fibonacci Quasicrystals}

\author{He-Guang Xu}
\affiliation{School of Physics, Dalian University of Technology, 116024 Dalian, China}

\author{Shujie Cheng}
\thanks{chengsj@zjnu.edu.cn}
\affiliation{Xingzhi College, Zhejiang Normal University, Lanxi 321100, China}
\affiliation{Department of Physics, Zhejiang Normal University, Jinhua 321004, China}

\date{\today}

\begin{abstract}

Extended and critical states are two common phenomena in Fibonacci quasicrystals. In this paper, 
we first reveal the difference between the extended phase and the critical phase in the extended-critical Fibonacci 
quasicrystal from the perspectives of quantum transport and Wigner distribution. The transport conductance indicates 
that the extended-critical transition resembles a metallic-insulating transition.
Moreover, the Wigner distributions show that the Wigner distribution of the extended wave function is localized 
in the momentum direction of the phase space, while that of the critical wave function is sub-extended in the momentum 
direction of the phase space. Based on the results of entanglement entropy, the extended-critical transition is a thermodynamic 
phase transition because it is accompanied by decreasing entropy. We engineer a 
quantum heat cycle engine with the extended-critical quasicrystal as the working medium, and find that 
there are rich working modes in the engine, such as quantum accelerator, quantum heater and quantum heat engine. 
 Importantly, the extended quasicrystals are more conducive to the realization of quantum heat engines, while the 
 critical quasicrystals are more conducive to the realization of quantum heaters. Our work is an important step 
 toward exploring the rich thermodynamic applications of Fibonacci quasicrystals.
\end{abstract}

\maketitle

\section{Introduction}
Anderson localization is a widely studied quantum phenomenon that exists in quantum systems 
with disorder or quasidisorder and explains how matter waves appear to lack ergodicity in such disordered \cite{ref1}
or quasidisordered environments \cite{ref2}. Its influence extends to a wide range of physical disciplines, including condensed 
matter physics \cite{ref3,ref4}, which helps elucidate the properties of electrons in disordered materials. In ultra-cold atomic 
systems, the study of Anderson localization has also been crucial in helping to understand the effects of complex disordered 
environments on particle behavior and driving the development of optical lattice experimental design and measurement 
techniques \cite{ref5,ref6,ref7,ref8,ref9,ref10,ref11,ref12,ref12_1,ref12_2,ref12_3}. In addition, the study of Anderson localization 
in photonic crystals will help understand the transmission behavior of light and promote the development of functional optical devices \cite{ref13,ref14,ref15,ref16,ref17,ref18,ref19}.

Under different physical mechanisms, Anderson localization presents different forms. For example, 
in the Fibonacci quasicrystal, when the strength of the quasidisordered potential exceeds to the critical value, 
all quantum states are localized \cite{ref2,ref6}. However, in the three-dimensional Anderson 
model \cite{ref3} or in systems with short- (long-) range hoppings \cite{ref20,ref21,ref22,ref23,ref24,ref25,ref26,ref26_1}, 
and generalized quasidisordered potentials \cite{ref27,ref28,ref29,ref30,ref31,ref32,ref33,ref34,ref35}, Anderson 
localization occurs only at certain energy levels, separated from the delocalized energy levels by the mobility edge. 
This makes the overall nature of the system appear to be in an intermediate phase that is neither fully extended nor 
fully localized. Based on the researches of mobility edge, the intermediate phase with extended-localized mobility 
edge has been found, which are used for the realizations of energy current 
rectification \cite{ref36} and superradiance light sources \cite{ref37}. In addition, it has been found that the electrical 
transport properties of the system can be significantly affected by properly designing and controlling the interaction 
between the system and the environment. This makes it possible to induce or annihilate the mobility edge through 
the interaction of the system with the environment, thereby altering the metallic or insulating properties of the system. 
For example, Liu et al. recently investigated the effect of bond dissipation on the Fibonacci quasicrystal with mobility edges and found 
that this dissipation mechanism can destroy mobility edges, causing the system to be fully extended  or fully localized 
in non-equilibrium steady-state \cite{ref38}. Here the extended-localized transition leads to the metallic-insulating transition of the system. 
Very recently, Longhi studied the influence of dephasing effects on Fibonacci quasicrytals with mobility edges and found that 
mobility edges do not disappear under dephasing effects. Meanwhile, the influence of dephasing on the off-diagonal Fibonacci 
quasicrystal with extended-critical phase transition was also studied, and the unexpected mobility edge induced by the 
dephasing effect was found \cite{ref39}. Although the wave function of the critical phase is sub-extended, it remains to be emphasized 
whether the critical quasicrystals behave more like metals or insulators in the transport measurements. Meanwhile, we note 
that the attentions are more payed to the localization physics of Fibonacci quasicrystals at zero temperature. In fact, it has 
been shown that Fibonacci quasicrystals with mobility edges interacting with external heat baths can be used to engineer 
quantum heat engines \cite{ref40}. However, for the Fibonacci quasicrystal without mobility edge, whether the interaction with 
the external heat source will trigger the quantum heat engine or richer applications remain to be explored.

 The paper is organized as follows. In Sec.~\ref{S2}, we introduce the Hamiltonian of the Fibonacci quasicrystal and study its electrical transport properties. 
 In Sec.~\ref{S3}, we reveal the localization properties from the aspects of Wigner distribution and entanglement entropy. 
 In Sec.~\ref{S4}, we explore the thermodynamic application of Fibonacci Quasicrystal. 
 A summary is made in Sec.~\ref{S5}.

\section{Model and Electrical transport properties}\label{S1}

\begin{figure}[htp]
		\centering
		\includegraphics[width=0.5\textwidth]{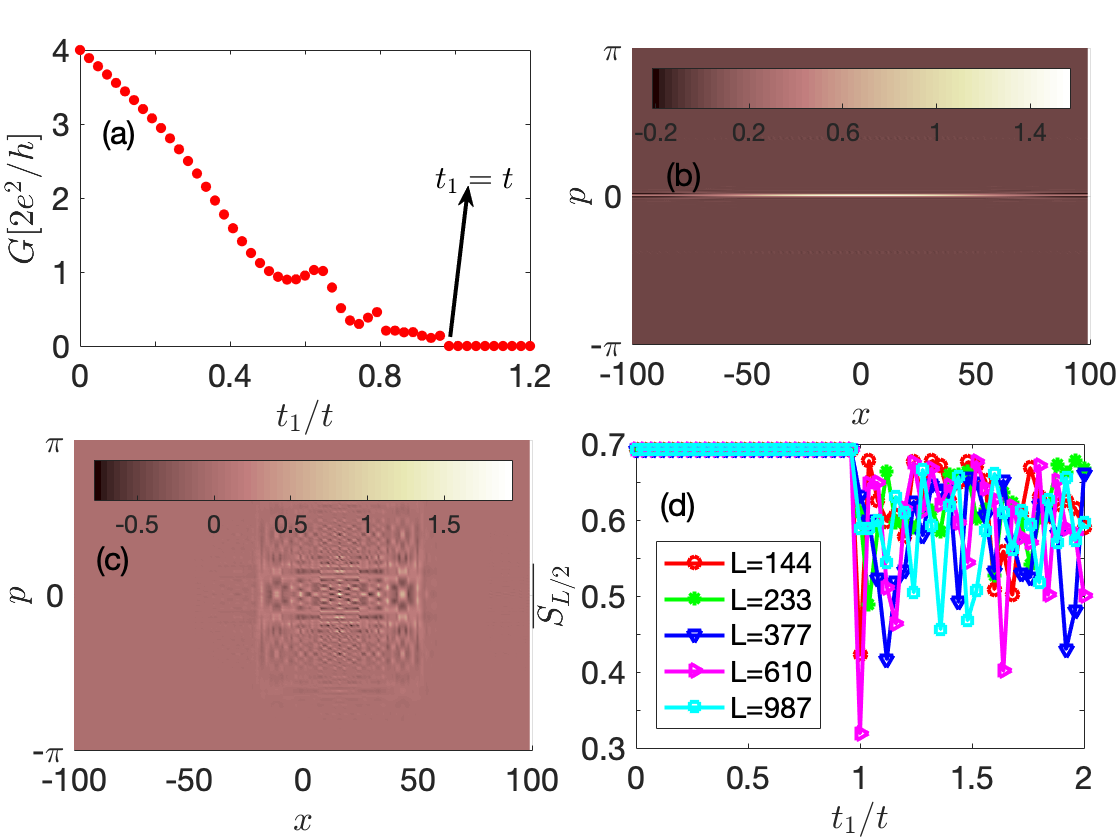}
		\caption{(Color online) (a) Transport conductance as a functions of $t_{1}$. The system size is 
		$L=987$ and the incoming energy is $E=0.01t$. (b) The Wigner distribution of the ground state 
		under $L=200$ and $t_{1}=0.2t$. (c)  The Wigner distribution of the ground state under $L=200$ 
		and $t_{1}=1.2t$. (d) The mean single-particle entanglement entropy $\overline{S_{1/2}}$ as 
		a function of $t_{1}$. 
		}
		\label{f1}
\end{figure}

We consider the off-diagonal Fibonacci quasicrystal described by the following 
Hamiltonian 
\begin{equation}
\hat{H}=-\sum_{n}\left(t_{n}\hat{c}^{\dag}_{n+1}\hat{c}_{n}+{\rm H.c.}\right),  
\end{equation}
in which $t_{n}=t+t_{1}\cos\left(2\pi\alpha n +\varphi\right)$ ( where $t$ is the unit the energy) denotes 
the incommensurate hopping amplitude (The commensuration is reflected in that the frequency $\alpha$ is 
the ratio of two adjacent infinite Fibonacci numbers.) between nearest neighbor sites $n$ and $(n+1)$ and $\hat{c}_{n}$, 
$\hat{c}^{\dag}_{n}$ are the annihilation and creation operators of the hard-core bosons at site $n$. The model presents 
extended-critical transition with the critical value $t_{c}=t$, and the phase diagram is shown in Fig.~\ref{f1}. Refs.~\cite{ref12_3,ref39,ref41,ref42,ref43} 
give the early proof of extended-critical transition in off-diagonal quasicrystals from perspectives of 
 (inverse) participation ratio, fractal dimension, and the spreading dynamics.
When $t_{1}<t_{c}$, the system is in the extended phase, and when $t_{1}>t_{c}$, the system is in the critical phase.  
In fact, the extended-critical transition can be understood by means of quantum transport measurement as well, which gives 
the transport conductance of the off-diagonal quasicrystal, the most direct characterization of the electrical transport 
properties of the systems. Although the wave function in the critical phase is sub-extended, it remains to be explored whether 
the critical quasicrystals behave more like metals or insulators in terms of their transport properties. 

Solving the conductance by quantum scattering is a feasible method in quantum transport theory \cite{ref44,ref45}, 
which is used to study the transport properties of electrons in quantum systems. In the following, we will 
briefly describe how to solve the transport conductance of this non-diagonal quasicrystal by the quantum scattering method.
By engineering a quantum transport device composed of a source, non-diagonal quasicrystals and a drain, and 
making incoming electrons with energy $E$ experience the process of transmission, reflection from source to drain, 
and scattering with a off-diagonal quasicrystals, we can obtain the quantum scattering matrix of the transport process, 
which is defined as 
\begin{equation}
S=\left(
\begin{array}{cc}
R & T' \\
T & R'
\end{array}
\right)
\end{equation}
in which $R$ and $R'$ denote the  reflection matrices while $T$ and $T'$ denote the transmission. 

The conductance (G) can be calculated by the transmission matrix $T$. In quantum transport theory, 
the conductance is related to the trace of $T$. Under an incoming energy $E$,  $G$ is defined as 
\begin{equation}
G=\frac{2e^2}{h}{\rm Tr} \left[TT^{\dag}\right], 
\end{equation}
where $e^2/h$ is the init of $G$ and the factor $2$ stands for the spin degeneracy. Considering the 
incoming energy $E=0.01t$, the conductance $G$ as a function of the hopping parameter $t_{1}$ is 
plotted in Fig.~\ref{f1}(a). As seen, when the off-diagonal quasicrystal undergoes the transition from 
the extended phase to the critical phase, there is an accompanying vanishing transport conductance.. 
It means that the extended-critical transition in the off-diagonal quasicrystal is more like a metallic-insulating transition. 
In this regard, in addition to the extended-localized quasicrystals, the extended-critical quasicrystal can be 
used as a quantum current limiter as well.

\section{Winger distribution and Entanglement entropy}\label{S2}
In addition to the quantum transport, the difference between the 
extended and critical phase can be readily seen from the Wigner distribution \cite{ref46,ref47,ref48,ref49}. The Wigner distribution 
function provides a way to describe quantum states in phase space. It maps the wave function of the 
quantum state to a real function of the phase space, so that the distribution of the quantum state 
can be intuitively understood in the phase space. For a given wave function $\psi(x)$ (where $x$ is the position), 
the corresponding Wigner distribution function  $W(x,p)$ ($p$ is the momentum) is defined as 
\begin{equation} 
W(x,p)=\frac{1}{2\pi\hbar}\int^{+\infty}_{-\infty}\psi(x+\frac{y}{2})\psi(x-\frac{y}{2})e^{-ipy/\hbar}dy, 
\end{equation} 
Considering the system size $L=200$, the ground-states (chosen as examples) $W(x,p)$ under 
$t_{1}=0.2t$ and $t_{1}=1.2t$ are plotted in Figs.~\ref{f1}(b) and \ref{f1}(c), respectively. The extended 
ground state corresponds to the extended distribution in the $x$ direction and the localized distribution 
in the $p$ direction in the phase space. The critical ground state corresponds to the sub-extended 
distribution in the $x$ and $p$ directions in the phase space. In addition, it can be seen that in some 
regions in the phase space, $W(x,p)$ are positive, while in others, the distributions are negative. Therefore, 
the results not only explain the extension-critical difference, but also reveal the property of Wigner 
distribution function as a quasi-probability distribution function. 

We reveal that the extended-critical transition is actually a thermodynamic phase transition. 
We here detect the thermodynamic property by the thermodynamic quantity, i.e., single particle entanglement entropy $S_{L/2}$, 
which is defined as 
\begin{equation}
S_{L/2}=-\sum_{j}\left[\epsilon_{j}\ln\epsilon_{j}+(1-\epsilon_{j})\ln(1-\epsilon_{j})\right],
\end{equation}
where the $\epsilon_{j}$ is the eigenvalues of the half-chain single-particle correlation matrix $C$,  
given by 
\begin{equation} 
C_{mn}=\braket{\psi|\hat{c}^{\dag}_{m}\hat{c}_{n}|\psi},
\end{equation}
with $\ket{\psi}$ the wave function. Taking different system size, the mean single-particle entanglement 
entropy $\overline{S_{L/2}}$ as functions of $t_{1}$ are plotted in Fig.~\ref{f1}(d). We can see that, when 
the hopping parameter $t_{1}$ exceeds the extended-critical transition point, there are decreasing 
single-particle entanglement entropy. In previous study, the realization of quantum heat engine is 
often accompanied by the high entropy state of the working medium \cite{ref49}. Thus, we speculate that here the 
extended quasicrystal is beneficial to the realization of the quantum heat engine. Notice that in 
the critical phase, $\overline{S_{L/2}}$ is only slightly lower than that in the extended phase, but it does 
not approach zero. Therefore, the underlying applications brought by the critical quasicrystal is an open question 
to be solved. Even in the extended phase, whether there are applications other than quantum heat engine 
is also worth exploring. 

\begin{figure}[htp]
		\centering
		\includegraphics[width=0.5\textwidth]{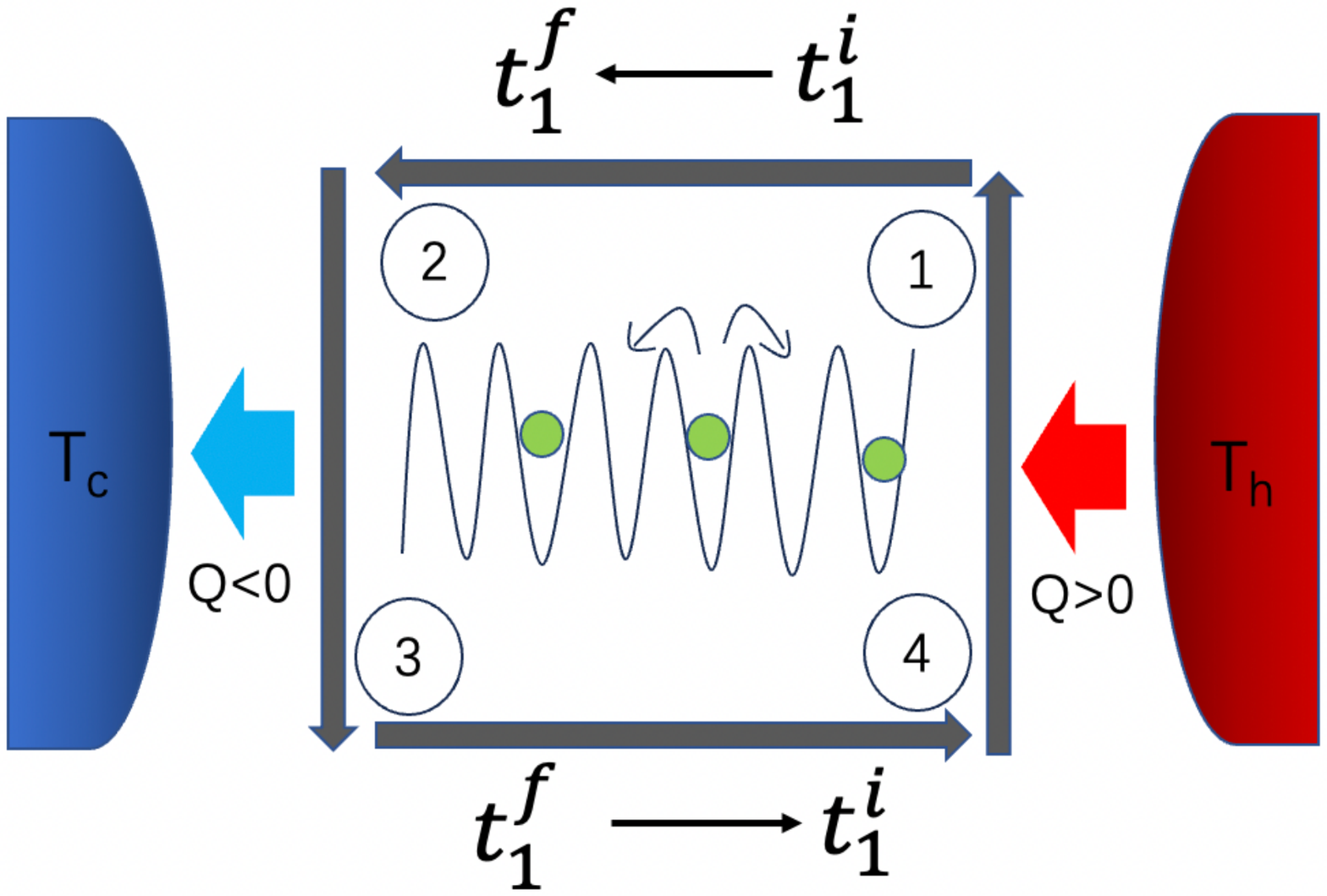}
		\caption{(Color online) Sketch of the four-strokes quantum heat cycle process. 
		$\rm{T_{h}}$ and $\rm{T_{c}}$ denote the high temperature and low temperature heat 
		sources, respectively. The working medium is the extended-critical quasicrystal. 
		$t^{f}_{1}$ and $t^{i}_{1}$ are the hopping parameters of corresponding Hamiltonians. 
		}
		\label{f2}
\end{figure}

\section{Quantum heat cycle} \label{S3}
We engineer a quantum heat cycle process using 
the extended-critical quasicrystal as the working medium, and the schematic diagram is displayed in 
Fig.~\ref{f2}. This heat cycle process can be viewed as the quantum analog \cite{ref49,ref50,ref51,ref52,ref53,ref54,
ref55,ref56,ref57,ref58,ref59,ref60,ref61} of the classical Otto cycle. The first ($\textcircled{4}\rightarrow\textcircled{1}$) and 
third ($\textcircled{2} \rightarrow \textcircled{3}$) strokes take place in heat contact with high ($T_{h}$) and 
low ($T_{c}$) temperature heat sources and run without external driving. The second ($\textcircled{1} \rightarrow \textcircled{2}$) 
and fourth ($\textcircled{3} \rightarrow \textcircled{4}$) strokes are adiabatic from a thermodynamic 
perspective (meaning they are isolated from heat), but they may not be adiabatic in a quantum-mechanical sense, 
as quantum transitions can take place during the evolution \cite{ref60}. 

In the first stroke, the working medium with the Hamiltonian $H(t^{i}_{1})$ will finally relax into the Gibbs 
state of thermal equilibrium, whose density matrix $\rho_{1}$ is given by $\rho_{1}=\frac{e^{-\beta_{h} H(t^{i}_{1})}}{Z_{1}}$ 
with $\beta_{h}=\frac{1}{k_{b}T_{h}}$ and partition function $Z_{1}={\rm Tr}(e^{-\beta_{h}H(t^{i}_{1})})$. 
Thus, in the thermal equilibrium, the energy of the system is $E_{1}={\rm Tr}\left[\rho_{1} H(t^{i}_{1})\right]$. During the second 
stroke, the Hamiltonian parameter is changed from $t^{i}_{1}$ to $t^{f}_{1}$ and only work is performed 
without any heat being exchanged. The Gibbs state of this stroke $\rho_{2}$ remain unchanged, i.e., $\rho_{2}=\rho_{1}$, 
but the energy $E_{2}$ becomes $E_{2}={\rm Tr}\left[\rho_{2}H(t^{f}_{1})\right]$. In the third stroke, the 
medium with $t_{1}=t^{f}_{1}$ is in contact with the source $\beta_{c}=\frac{1}{k_{b}T_{c}}$. Hence, the Gibbs 
state $\rho_{3}$ gives $\rho_{3}=e^{-\beta_{c}H(t^{f}_{1})}/Z_{2}$ with partition function $Z_{2}={\rm Tr}\left[e^{-\beta_{c}H(t^{f}_{1})}\right]$. 
Then, the energy of the medium $E_{3}$ is $E_{3}={\rm Tr}\left[\rho_{3}H(t^{f}_{1})\right]$. The fourth stroke 
can be viewed as the quantum annealing process. In this process, the hopping parameter in the medium is changed 
from $t^{f}_{1}$ back to $t^{i}_{1}$. However, the Gibbs state $\rho_{4}$ remains $\rho_{4}=\rho_{3}$. Therefore, 
the energy of medium becomes $E_{4}={\rm Tr}\left[\rho_{4}H(t^{i}_{1})\right]$. After the working medium has 
experienced this complete cycle process, we can obtain the absorbed heat $Q_{h}$ from the $T_{h}$ source, 
and the released heat $Q_{c}$ to the $T_{c}$ source, as well as the net work $W$ of the working medium $W=Q_{h}+Q_{c}$. 
Importantly, this heat cycle process obeys the Clausius inequality, which is one of the cornerstones of thermodynamics. 
According to different values of $Q_{1}$, $Q_{2}$, and $W$, the engine with extended-critical quasicrystal as the 
working medium will present different modes \cite{ref59,ref60}: (1) {\it Heat engine}: $Q_{h}>0$, $Q_{c}<0$, and $W>0$; (2) {\it Refrigerator}: $Q_{h}<0$, 
$Q_{c}>0$, and $W<0$; (3) {\it Heater}: $Q_{h}<0$, $Q_{c}<0$, and $W<0$; (4) {\it Accelerator}: $Q_{h}>0$, 
$Q_{c}<0$, and $W<0$. 

\begin{figure}[htp]
		\centering
		\includegraphics[width=0.5\textwidth]{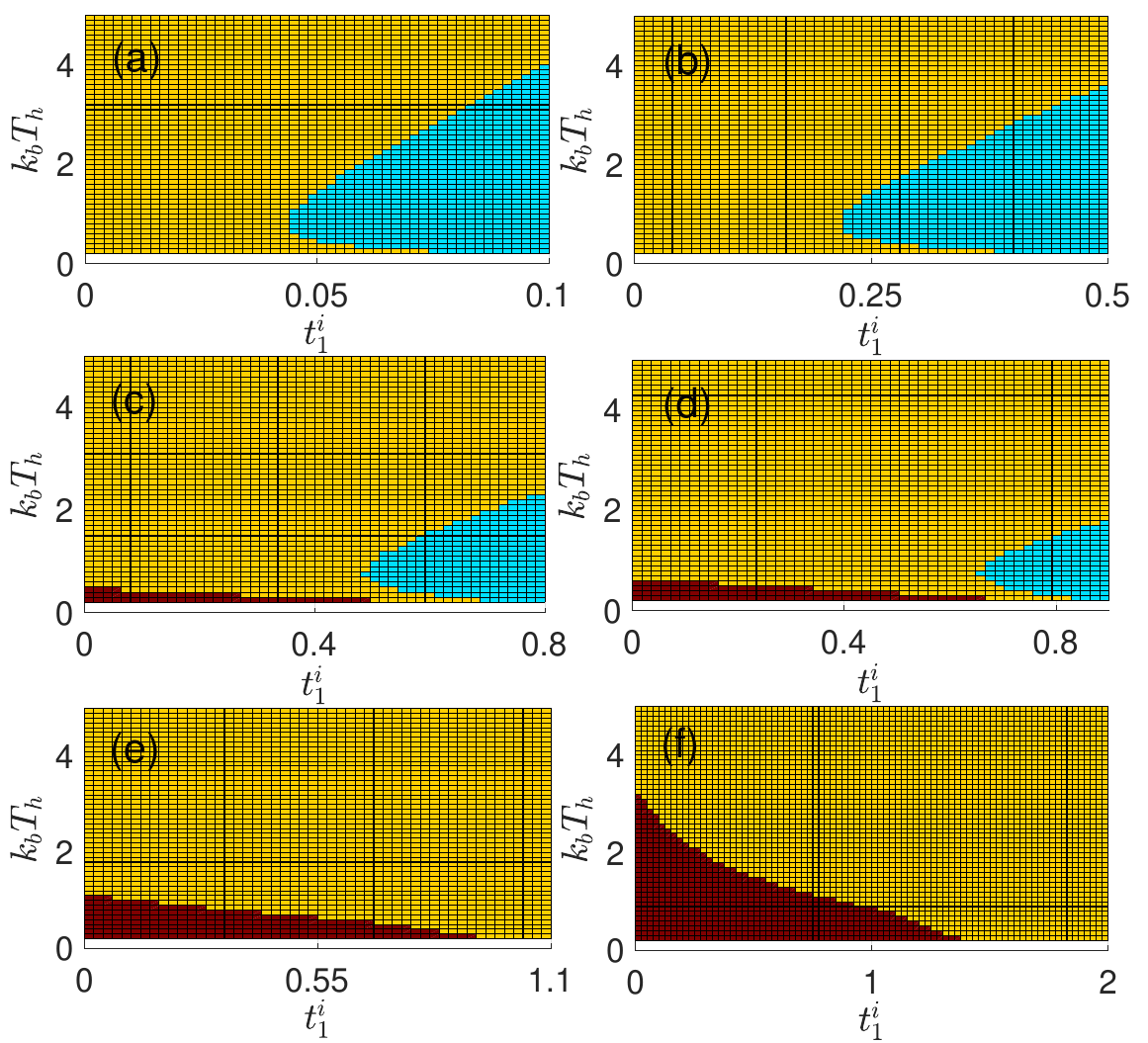}
		\caption{(Color online) Modes of the four-stroke engine in the $k_{b}T_{h}-t^{i}_{1}$ parameter 
		plane. (a) $t^{f}_{1}=0.1t$.  (b) $t^{f}_{1}=0.5t$.  (c) $t^{f}_{1}=0.8t$.  (d) $t^{f}_{1}=0.9t$. 
		 (e) $t^{f}_{1}=1.1t$.  (f) $t^{f}_{1}=2t$. The blue regions denote the {\it Heat engine}. The 
		 yellow regions denote the {\it Accelerator} and the brown regions denote the {\it Heater}. 
		Other parameters are $L=233$ and $k_{b}T_{c}=0.1t$.  
		}
		\label{f3}
\end{figure}

Taking the system size $L=233$ and $k_{b}T_{c}=0.1t$, after analyzing the values of $Q_{1}$, $Q_{2}$, 
and $W$, the corresponding modes of the four-stroke engine with $t^{f}_{1}$ equaling to $0.1t$, $0.5t$, 
$0.8t$, $0.9t$, $1.1t$, and $2t$ are plotted  in Figs.~\ref{f3}(a)-\ref{f3}(f), respectively. We can see that 
this engine with extended-critical quasicrystal as the working medium has rich working modes. The blue regions 
denote the {\it Heat engine}. The yellow regions denote the {\it Accelerator} and the brown regions 
denote the {\it Heater}. We note that when $t^{i}_{1}$ and $t^{f}_{1}$ are less than the critical value $t_{c}$, 
there is the mode of {\it Heat engine}, as the blue regions show. It means that the extended 
quasicrystal are more conducive to the design of quantum heat engine. When $t^{f}_{1}$ gradually gets larger, 
and exceeds the critical values, we can see that the regions characterizing the mode of the {\it Heater} get larger 
as well. It means that the critical quasicrystal are more conductive to the realization of the quantum heater. Besides, 
it is evident that there are extensive parameter regions that characterize the mode of the {\it Accelerator}, regardless 
of whether the quasicrystal is in the extended phase or the critical phase. It implies that the extended 
and critical quasicrystals are both conductive to the realization of the quantum accelerator.

In addition, the results presented in Fig.~\ref{f3} provide strategies for regulating the working mode of 
the four-stroke engine as well. The transition between different working modes can be achieved by adjusting $t^{f}_{1}$, $t^{i}_{1}$, 
and $k_{b}T_{h}$. For instance, when $t^{f}_{1}$ is far less than $t_{c}$ (see Figs.~\ref{f3}(a) and \ref{f3}(b)), 
one can adjust $t^{i}_{1}$ or $k_{b}T_{h}$ to make the working mode of the engine change from the {\it Accelerator} to 
the {\it Heat engine}, or from the {\it Heat engine} to the {\it Accelerator}, respectively. When $t^{f}_{1}$ is close to 
but still less than $t_{c}$, there are three working modes when $k_{b}T_{h}$ is small. Therefore, the working mode of the 
engine will change from the {\it Heater} to the {\it Accelerator} and then to the {\it Heat engine}. When $t^{f}_{1}$ is larger than 
$t_{c}$, the engine can be switched between heater and accelerator by adjusting $t^{i}_{1}$ or $k_{b}T_{h}$. 

\section{Summary}\label{S4}
 In summary, we have revealed that the extended-critical Fibonacci quasicrystal 
presents metallic-insulating transition from the perspective of quantum transport. Besides, there is a quantum-classical 
correspondence between the wave function and the Wigner distribution in phase space. Specifically, the extended 
wave function corresponds to the localized Wigner distribution in the momentum direction in the phase space, while 
the critical wave function corresponds to the sub-extended distribution in the momentum direction. Importantly, the 
single-particle entanglement entropy shows that the extended-critical transition is actually a thermodynamic transition, 
due to the decrease of the entanglement entropy when the  Fibonacci quasicrystal goes from the extended 
phase to the critical phase. The thermodynamic transition leads to rich applications of the quantum Otto heat cycle engine 
with extended-critical Fibonacci quasicrystal being the working medium, including quantum heat engine, quantum accelerator, 
and the quantum heater. Interestingly, the extended Fibonacci quasicrystal are more conductive to be employed to engineer 
the quantum heat engine, while the critical Fibonacci quasicrystal are more conductive to be employed to engineer the quantum 
heater. Both the extended and critical phase are conductive to the realization of quantum accelerator. By adjusting the 
quasicrystal parameters and the temperature of the heat source, we can realize the transition between different 
working modes.

 This work is supported by the start-up fund from Xingzhi college, Zhejiang 
 Normal University.

\bibliography{references}

\end{document}